\documentclass[prd, showpacs, twocolumn,superscriptaddress]{revtex4}

\usepackage{graphicx}

\usepackage[normalem]{ulem}
\usepackage[usenames]{color}

\def\lsim{\raise0.3ex\hbox{$\;<$\kern-0.75em\raise-1.1ex
\hbox{$\sim\;$}}}
\def\gsim{\raise0.3ex\hbox{$\;>$\kern-0.75em\raise-1.1ex
\hbox{$\sim\;$}}}

\begin{document}
\title{Bulk Neutrinos as an Alternative Cause of the Gallium and Reactor
  Anti-neutrino Anomalies}

\author{P.~A.~N.~Machado}
\email{accioly@fma.if.usp.br}
\affiliation{
Instituto de F\'{\i}sica, Universidade de S\~ao Paulo, 
 C.\ P.\ 66.318, 05315-970 S\~ao Paulo, Brazil}
\affiliation{Institut de Physique Th\'eorique, CEA-Saclay, 91191 Gif-sur-Yvette, France}
\author{H.~Nunokawa}
\email{nunokawa@puc-rio.br} 
\affiliation{Departamento de F\'{\i}sica,
  Pontif{\'\i}cia Universidade Cat{\'o}lica do Rio de Janeiro,
  C. P. 38071, 22452-970, Rio de Janeiro, Brazil}
\author{F.~A.~ Pereira dos Santos}\email{fabio.alex@fis.puc-rio.br}
\affiliation{Departamento de F\'{\i}sica,
  Pontif{\'\i}cia Universidade Cat{\'o}lica do Rio de Janeiro,
  C. P. 38071, 22452-970, Rio de Janeiro, Brazil}
\author{R.~Zukanovich Funchal} \email{zukanov@if.usp.br} \affiliation{
  Instituto de F\'{\i}sica, Universidade de S\~ao Paulo,
  C.\ P.\ 66.318, 05315-970 S\~ao Paulo, Brazil}
\vglue 1.4cm

\begin{abstract}
We consider an alternative explanation for the deficit of $\nu_e$ in
Ga solar neutrino calibration experiments and of the $\bar\nu_e$ in
short baseline reactor experiments by a model where
neutrinos can oscillate into sterile Kaluza-Klein modes that can
propagate in compactified sub-micrometer flat extra dimensions.  We
have analyzed the results of the gallium radioactive source
experiments and 19 reactor experiments with baseline shorter than 100
m, and showed that these data can be fitted into this scenario.  The
values of the lightest neutrino mass and of the size of the largest
extra dimension that are compatible with these experiments are mostly
not excluded by other neutrino oscillation experiments.
\end{abstract} 

\maketitle

\section{Introduction}
\label{sec:intro}

We have been living exceptional times in neutrino physics. 
Neutrino mixings and masses have been  substantiated by
 a plethora of oscillation experiments which favor 
the standard three  flavor mixing scheme. 
Solar~\cite{solar} and atmospheric~\cite{atmos} neutrino 
experiments have established two fairly large mixing angles 
and two distinct mass squared differences which today are  
rather precisely determined by reactor~\cite{kl} and accelerator 
experiments~\cite{k2k,minos}. Recently T2K~\cite{T2K} has 
announced that their data provides indication of a non-zero, and 
perhaps far from negligible, value of $\theta_{13}$, 
supported also by 
MINOS~\cite{Adamson:2011qu}, opening the auspicious 
possibility to access CP violation in the leptonic sector by 
current or near future experiments.

While all the neutrino data mentioned above can be fitted 
very well into the standard picture of the three flavor neutrino
scheme, there have been some data~\cite{LSND,MiniBooNE} 
which are not consistent with such a picture. 
First, the LSND~\cite{LSND} experiment has observed an 
excess of $\bar{\nu}_e$ events in the $\bar{\nu}_\mu \to \bar{\nu}_e$
mode, which seemed to be supported by 
MiniBooNE data~\cite{MiniBooNE},
indicating the presence of at least one species of the so-called 
sterile neutrinos. 
These neutrinos would have to be separated from the active neutrinos 
by a mass squared difference of $\sim$ eV$^2$. 
Let us call this the LSND/MiniBooNE anomaly.

Likewise, calibrations of the gallium 
radiochemical solar neutrino detectors 
of GALLEX~\cite{gallex} and SAGE~\cite{sage} experiments
performed using intense portable neutrino radioactive 
sources, $^{51}$Cr by GALLEX and SAGE, and $^{37}$Ar by SAGE, 
observed some deficit of $\nu_e$ compared 
to what was expected, giving rise to the so-called 
{\em gallium anomaly}.
The mean value of the ratios of the measured over predicted 
rates is $0.86 \pm 0.05$ which is smaller than unity by 
about 2.7 $\sigma$~\cite{giunti2}.
This can also be explained by 
oscillation into sterile neutrinos with 
the similar mass squared difference which explains 
the LSND/MiniBooNE anomaly.  

More recently, a re-evaluation of the reactor $\bar \nu_e$
flux~\cite{Mueller:2011nm,Huber:2011wv} performed in order to prepare 
for the Double Chooz 
reactor experiment~\cite{Ardellier:2004ui}
resulted in an increase in the flux of 3.5\%. 
While this increase has essentially no impact on
the results of long baseline experiments such as KamLAND,
it induces an average deficit of 5.7\% 
in the observed event rates for short baseline ($ < 100$ m) reactor 
neutrino experiments
leading to the 98.6\% CL deviation from unity,
which has been referred to as 
the {\em reactor antineutrino anomaly}~\cite{reactor-anomaly}. 

It was shown in Ref.~\cite{reactor-anomaly} that these three anomalies
can be explained by a phenomenological $3+1$ model, where the
oscillation scheme involves the three active neutrinos and one
additional species of sterile neutrino. In
Refs.~\cite{Kopp:2011qd,Giunti:2011gz} it was performed a global fit
of the short baseline experiments (but without Ga data) with sterile
neutrinos and it was concluded that data can be fitted significantly
better in a 3+2 model.

In the interim, however, the LSND/MiniBooNE anomaly has diminished 
substantially. A more recent MiniBooNE result, based on the
$8.58 \times 10^{20}$ POT, reduced the significance of the
$\bar{\nu}_\mu \to \bar{\nu}_e$ excess 
to 0.84 $\sigma$~\cite{MiniBooNE-Nufact11} and very recently the HARP-CDP
Group~\cite{Bolshakova:2011hr} presented new data on pion production that also
decreased the significance of the LSND excess from 
3.8 to 2.9 $\sigma$. 

In this paper, we will show that the two anomalies seen in the
disappearance channels, the gallium and the antineutrino reactor ones,
can be accommodated in a scenario where three right
handed neutrinos propagate in a higher dimensional bulk, including a
large compactified flat extra dimension~\cite{ADD}, and all Standard
Model particles are confined to a 4-dimensional brane. The 3 bulk
fermions have Yukawa coupling with the Higgs and the brane neutrinos
leading to small Dirac neutrino masses and mixings among active
species and sterile Kaluza-Klein
modes~\cite{Dienes:1998sb,ArkaniHamed:1998vp,Dvali:1999cn,
  Barbieri-et-al,Mohapatra-et-al,Davoudiasl:2002fq}.

It is important to emphasize that the model presented here is
significantly different from the phenomenological models studied in
Refs. \cite{reactor-anomaly,Kopp:2011qd,Giunti:2011gz}. 
In general, a 3+$n$ phenomenological model assumes that 
the three active neutrinos can mix with $n$ sterile species 
which implies that, in addition to the 2+$n$ mass squared differences,  
the (3+$n$)(2+$n$)/2 mixing angles and (2+$n$)(1+$n$)/2 
phases are free parameters relevant for oscillation physics.
Therefore, the number of relevant parameters for the 3+$n$ model
is significantly larger than that of the standard three
flavor scheme. 
We, however, note that in the phenomenological approach, 
usually, the numbers of free parameters used in the fit 
are reduced to simplfy the analysis in thses models, 
as done in Refs.~\cite{reactor-anomaly,Kopp:2011qd,Giunti:2011gz}.

On the other hand, besides providing an explanation for 
the smallness of neutrino masses
\cite{Dienes:1998sb}, the free parameters of the LED model described
here that can have some impact on oscillation physics are the 3 mixing
angles, one $CP$ phase, the radius of the largest extra dimension and
the neutrino mass scale. The mixing between the active neutrinos and
the KK sterile modes is completely determined by these parameters. 
So, despite being (innately) conceptually more elaborated than the
phenomenological $3+n$ models, the LED model considered in this work 
is intrinsically much more constrained as a model, 
once it involves less free parameters \cite{footnote1}.

This alternative explanation is consistent with the results of the
current terrestrial experiments such as CHOOZ~\cite{chooz},
KamLAND~\cite{kl} and MINOS~\cite{minos} limits and seem to be
consistent with solar~\cite{solar} and atmospheric~\cite{atmos}
oscillation as discussed in Ref.~\cite{mnz2010}.  However, the
$\bar{\nu}_e$ excess observed in the LSND and MiniBooNE experiments
can not be explained by the scenario addressed here, and therefore we
do not consider them in this work.

\section{Neutrino Oscillations with a Large Extra Dimension}

\begin{figure*}[!t]
\begin{center}
\hglue 0.2cm
\includegraphics[scale=0.66]{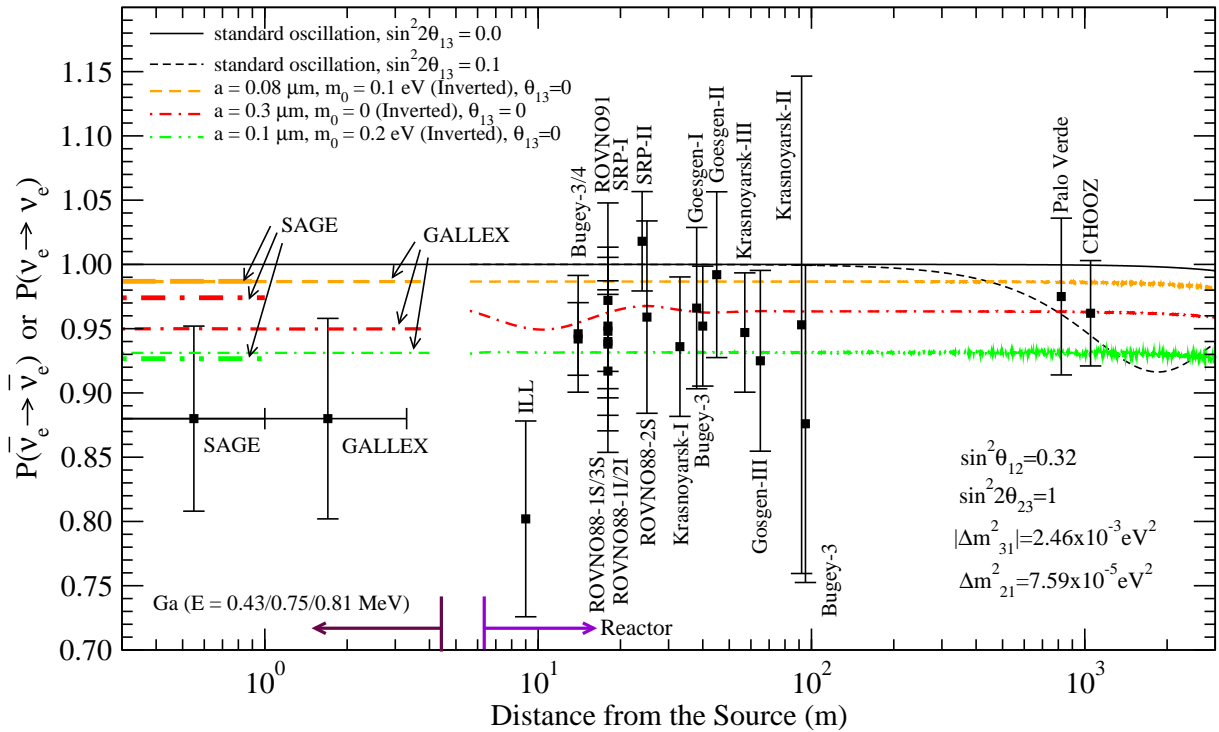}
\end{center}
\vglue -3.0cm
\caption{Survival probability as a function of the distance from 
the $\nu_e$ ($\bar \nu_e$) source averaged over the detector position 
(reactor energy spectrum).  To illustrate how LED, in principle,  
can explain the short baseline anomalies we show 
this probability for (a) standard
oscillation with $\sin^2 2\theta_{13}=0.0$ (black continuous line) and 
$\sin^2 2\theta_{13}=0.1$ (black dashed line) and (b) LED for IH with 
$\theta_{13}=0$ and some values of the LED parameters:  
$a=0.3$ $\mu$m, $m_0=0$  (red lines), 
$a=1.0$ $\mu$m, $m_0=0.2$ eV  (green lines), 
$a=0.08$ $\mu$m, $m_0=0.1$ eV  (orange lines).
The continuous, dashed and dotted lines refer to 
the reactor, GALLEX and SAGE experiment, respectively. 
We also show the average experimental deficits 
for the source test experiments SAGE and GALLEX, 
as well as for the reactor experiments ILL, Bugey-3, 
Bugey-3/4, ROVNO88-1S/3S, ROVNO91, 
ROVNO88-1I/2I, ROVNO88-2S, SRP-I, SRP-II, G\"{o}sgen-I, G\"{o}sgen-II, G\"{o}sgen-III, 
Krasnoyarsky-I, Krasnoyarsky-II, Krasnoyarsky-III, Palo Verde and CHOOZ 
where the reactor data were taken from the Table II of 
Ref.~\cite{reactor-anomaly}.}
\label{fig:prob-ee}
\end{figure*}

The large extra dimension (LED) picture we will consider here is 
the one described in 
Refs.~\cite{Davoudiasl:2002fq,mnz2010}. 
There the 3 standard model (SM) left-handed flavor neutrinos 
fields $\nu_{\alpha}$
($\alpha=e,\mu,\tau$), as well as all the other SM fields, are confined 
to propagate in a 4-dimensional brane, while 3
SM singlet fermion fields can propagate in a higher
dimensional bulk, with at least two compactified extra dimensions. 
To retain simplicity, we will assume that one of these extra dimensions, 
compactified on a circle of radius $a \lsim 1\ \mu{\rm m}$~\cite{mnz2010}, 
is however much larger than the size of the others so that in practice 
a 5-dimensional treatment is enough.

The 3 bulk fermions have Yukawa couplings with the SM Higgs and
the brane neutrinos ultimately leading to flavor oscillations
driven by Dirac masses, $m_i$ ($i$=1,2 and 3), 
and Kaluza-Klein (KK) masses $m^{\text{KK}}_n$ ($n=1,2,...$),  
and mixings among active species and sterile modes.
In this case the $\nu_e$ (same as $\bar \nu_e$ due to CPT conservation) 
survival probability in vacuum can be written 
as~\cite{Davoudiasl:2002fq,mnz2010}
\begin{equation}
P(\nu_e \to \nu_e;L,E) = \vert {\cal{A}}_{\nu_e\to \nu_e} (L,E)\vert^2 \, ,
\end{equation}
where the amplitude is given by
\begin{equation}
{\cal{A}}_{\nu_e \to \nu_e} (L,E) =  \sum_{i=1}^{3} \vert U_{ei}\vert^2 A_i, 
\end{equation}
where $A_i$ is given by, assuming $m_ia \ll 1$ and ignoring the terms
of order $(m_ia)^3$ and higher in the amplitude as well as $(m_ia)^2$
and higher in the phase,
\begin{eqnarray}
\hskip -1cm
A_i   & \approx  & (1 - \frac{\pi^2}{6} m_i^2 a^2)^2 \exp \left(i \frac{m_i^2 L}{2E}\right) \nonumber \\
\hskip -1cm
& + & \sum_{n=1}^{\infty} 2 \left(\frac{m_i}{{m_n^{\text{KK}}}}\right)^2
\exp \left[i \frac{(2\,m_i^2+{m_n^{\text{KK}}}^2) L}{2E}\right].  
\label{eq:amplitude}
\end{eqnarray}
Here $U_{ei}$ are the elements of the first row of 
the usual Maki-Sakata-Nakagawa neutrino mixing matrix 
(we use the standard parameterization found in Ref.~\cite{Nakamura:2010zzi}),
$E$ is the neutrino energy, $L$ is the baseline distance, 
$m^{\text{KK}}_n =n/a$  is the mass of the $n$-th KK mode.

This survival probability depends on the neutrino
mass hierarchy, for normal hierarchy (NH) we have $m_3 > m_2 > m_1=m_0$ and
inverted hierarchy (IH) we have $m_2> m_1 > m_3=m_0$. 
Clearly, as $m_0$ 
increases the differences between the hierarchies 
fade away and the masses become degenerate.
So besides the usual oscillation parameters 
$\Delta m^2_{32}=\vert m^2_3-m^2_2\vert$,
$\Delta m^2_{21}= m^2_2-m^2_1$, $\sin^2 \theta_{12}$, $\sin^2\theta_{13}$,
which are basically fixed by the data from 
the current oscillation experiments,
LED oscillations will be also driven by
$a$ and $m_0$ which also have been 
constrained by experimental data~\cite{Davoudiasl:2002fq,mnz2010}.
Throughout this work, even in the presence of 
LED, we consider, to a good approximation, the following 
true (input) values of the standard oscillation parameters 
determined by the three flavor analysis of experimental data:
$\Delta m_{21}^2 = 7.6 \times 10^{-5}$ eV$^2$, 
$\sin^2 2\theta_{12}$ = 0.31, $|\Delta m_{31}^2| = 2.4 \times 10^{-3}$ eV$^2$.

If $a \lsim 1$ $\mu$m, the LED effect at short baselines is simply to promote 
$\nu_e \to \nu_n^{\text{KK}}$, 
converting part of the 
$\nu_e$ 
 signal into KK modes, producing 
a nearly energy independent 
depletion of the $\nu_e$ 
rates, and the same applies to antineutrinos. To illustrate this
we show in Fig.~\ref{fig:prob-ee} the survival probability for a few
sets of LED parameters as well as the radioactive source test
experiments and reactor rates.

How can one understand these results?  One can easily show that for
the short-baseline experiments, to leading order, the averaged 
surviving probability with the LED effect is 
\begin{equation}
\langle P(\nu_e \to \nu_e) \rangle \approx \left[\, \sum_{i=1}^{3}
  \vert U_{ei} \vert^2 \, 
\left(1- \frac{\pi^2 \, m_i^2 \, a^2 }{6}\right)^2 \right]^2.
\end{equation}
Therefore, if $a=0.3$ $\mu$m $\approx 3/2$ eV$^{-1}$, $m_3=m_0=0$,
$m_1 \approx m_2 \sim 0.05$ eV or if $a=0.1$ $\mu$m $\approx 1/2$
eV$^{-1}$, $m_3=m_0=0.2$ eV $\approx m_1 \approx m_2$ or if $a=0.08$
$\mu$m $\approx 2/5$ eV$^{-1}$, $m_3=m_0=0.1$ eV $\approx m_1 \approx
m_2$, the survival probability can be estimated as $\sim 1-2\, \pi^2\,
a^2 m_{2}^2/3$, given respectively, $\sim$ 0.96, 0.93, 0.99.

\section{Analysis Results}

\subsection{Gallium Radioactive Source Experiments}
\label{sec:ga}

Let us first look at the gallium anomaly.  The radiochemical solar
neutrino experiments GALLEX and SAGE have been calibrated with
monoenergetic $\nu_e$'s from intense radioactive sources, 
which are captured by the reaction,
\begin{equation}
\nu_e + ^{71}{\rm Ga} \to ^{71}{\rm Ge} + e^-
\label{Eq:Ga-reaction}.
\end{equation}
GALLEX collaboration published the results of their measurements with two
$^{51}{\rm Cr}$ sources~\cite{gallex}. SAGE collaboration performed
similar measurements with $^{51}{\rm Cr}$ and also with $^{37}{\rm Ar}$
sources~\cite{sage}.

They presented their results in terms of a ratio, $R$, of the measured
$^{71}{\rm Ge}$ event rate over the predicted one 
using the predicted cross section for the reaction (\ref{Eq:Ga-reaction})
estimated in Ref.~\cite{bahcall}, including errors. 
All the measured ratios are below unity, 
\begin{eqnarray}
R^{\rm G}_{\rm Cr1} &=& 0.95 \pm 0.11, \\
R^{G}_{Cr2} &=& 0.81 \pm 0.11,
\end{eqnarray}
for GALLEX~\cite{gallex} 
we used the values based on the pulse shape analysis obtained by Kaether  
{\it et al.} in Ref.~\cite{gallex} and 
 \begin{eqnarray}
R^{\rm S}_{\rm Cr} &=& 0.95 \pm 0.12, \\
R^{\rm S}_{\rm Ar} &=& 0.79 \pm 0.09,
\end{eqnarray}
for SAGE~\cite{sage}.

An analysis of these results in terms of oscillation of $\nu_e$ into
sterile neutrinos was performed in Refs.~\cite{giunti1,giunti2}.

We have done an analysis similar to the one described in Ref.~\cite{giunti1}.
Tab.~\ref{tab:ga-sa} shows the data needed to perform our calculation.
We have computed the theoretical value of the ratio $R$ for LED as 

\begin{equation}
R = \displaystyle \frac{\int dV \, L^{-2}  \; P(\nu_e \to \nu_e; L, E)}
{\int dV \, L^{-2}}
\label{eq:r}
\end{equation}
by Monte Carlo integration. In fact, there are four different $\nu_e$ lines 
emitted by the $^{51}$Cr source, two with energy around 750 keV and 
cross section $\sim 61 \times 10^{-46}$ cm$^{2}$ (90\%) and two 
with energy around 430 keV and cross section  $\sim 27 \times 10^{-46}$ 
cm$^{2}$ (10\%) and two  different $\nu_e$ lines 
emitted by the $^{37}$Ar source, both with energy around 810 keV.
We have verified that taking into account these different contributions 
in the calculation of $R$ as in Ref.~\cite{giunti2} does not affect our 
final result, so for the purpose of this paper it is enough to use 
Eq.~(\ref{eq:r}). 

\begin{figure*}[!t]
\begin{center}
\includegraphics[scale=0.52]{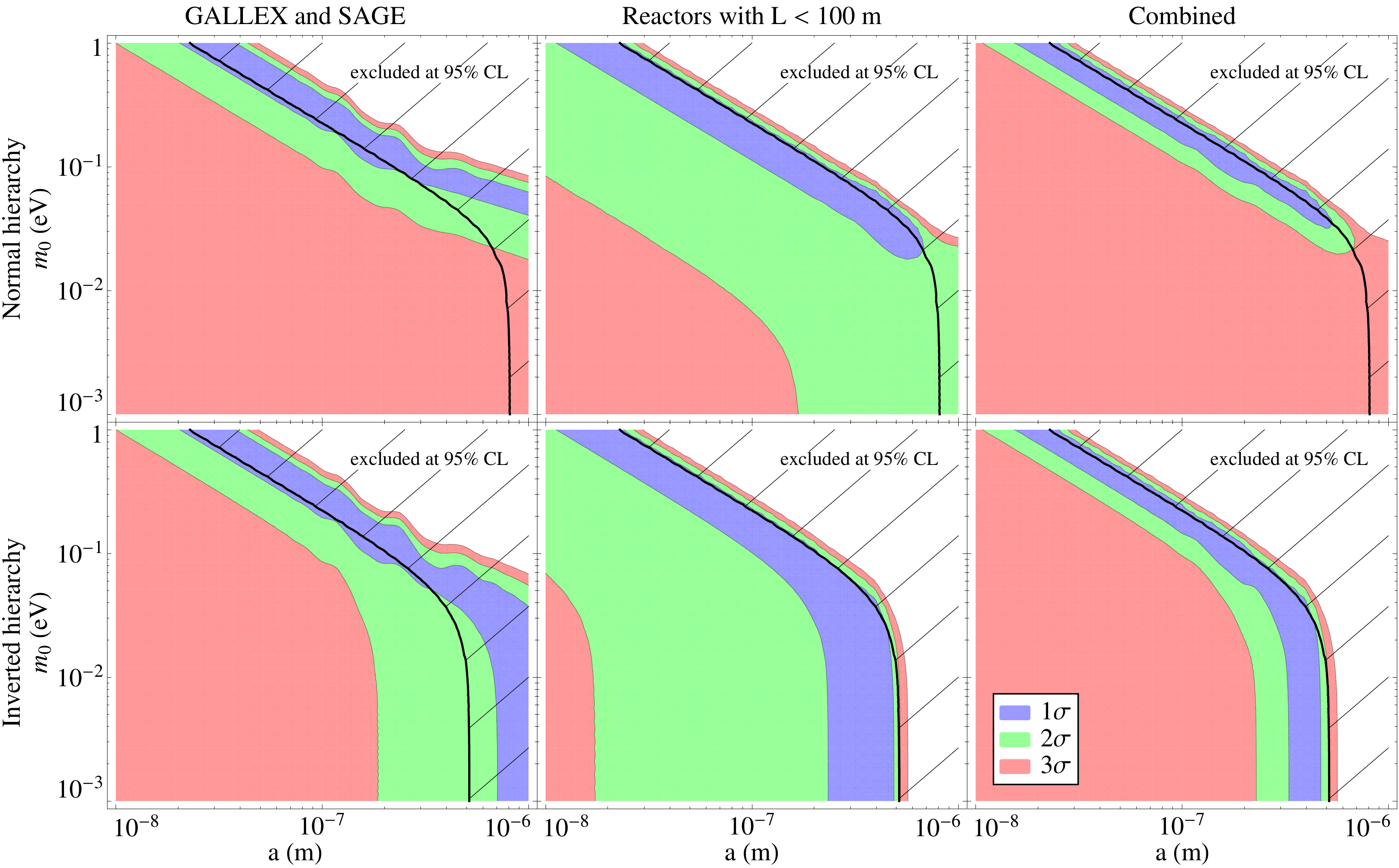}
\end{center}
\vglue -0.6cm
\caption{ Regions in the plane $m_0$ versus $a$ at 68\% , 95\% and
  99.73\% CL (that is 1, 2, and 3 $\sigma$) allowed by GALLEX and SAGE
  source calibration experiments (left panels), by short baseline
  reactor data (middle panels) and by the combined case of these two
  data set (right panels).  For each case, the upper (lower) panel
  correspond to the normal (inverted) hierarchy. The hatched areas
  correspond to the 95\% CL limits from terrestrial oscillation
  experiments derived in Ref.~\cite{mnz2010}.}
\label{fig:allowed-regions}
\end{figure*}

\begin{table}[!t]
\vglue 0.5 cm
\begin{centering}
\begin{tabular}{c|cc|cc}
\hline
& \multicolumn{2}{c|}{GALLEX} & \multicolumn{2}{c}{SAGE} \\
\hline
& Cr1 & Cr2 & Cr & Ar \\
\hline
E (keV) & \multicolumn{2}{c|}{750} & 750 & 811 \\
\hline
r (m) & \multicolumn{2}{c|}{1.9} & \multicolumn{2}{c}{0.7} \\
\hline
h (m) & \multicolumn{2}{c|}{5.0} & \multicolumn{2}{c}{1.47} \\
\hline
source position (m) & 2.7 & 2.38 & \multicolumn{2}{c}{0.72} \\
\hline
\end{tabular}
\par\end{centering}
\caption{For the GALLEX and SAGE source experiments we give the $\nu_e$ 
  energy (E) of the primary $\nu_e$ line emitted by the source, 
  the radius (r) and height (h) of the
  cylindrical detector volumes and the position of the sources in
  terms of height from the base of the detectors. The sources were
  placed along the axes of the detectors.} 
\label{tab:ga-sa}
\end{table}
We have performed a $\chi^2$ analysis of the data and found a region
allowed for the LED parameters $m_0$ and $a$ that fit well the for
data points, $\chi^2_{\min}/\rm dof=1.81/2=0.905$.  In the left panels
of Fig.~\ref{fig:allowed-regions}, the allowed regions are shown for
NH (upper panel) and IH (lower panel).  For the purpose of comparison
we also indicated in all the panels, by a solid curve, the region
excluded by KamLAND, CHOOZ and MINOS obtained in Ref.~\cite{mnz2010}.
We note that the 1 $\sigma$ allowed region is basically excluded by other
experiments but there are still large 2 and 3 $\sigma$ regions 
which are not in conflict with them. 
In fact, from Fig.~\ref{fig:prob-ee} one can expect that there could be
some ``tension'' between the Ga and  reactor data as the former 
prefer somewhat stronger reduction than the latter.

\subsection{Short Baseline Reactor Neutrino Experiments}
\label{sec:reac}

Using the new reactor antineutrino flux
calculations~\cite{Mueller:2011nm,Huber:2011wv} the ratio between the
number of $\bar \nu_e$ observed and theoretically predicted for all
short baseline reactor experiments has decreased by
5.7\%~\cite{reactor-anomaly}.

We have simulated the expected rates of the following 19 reactor
experiments with baselines shorter than $100$ m:
Bugey-3-I/III~\cite{bugey3} at 15, 40 and 90 m, of
Bugey-4~\cite{bugey4} at 15 m, of ILL~\cite{ill} at 9 m, of
G\"{o}sgen-I/III~\cite{gosgen} at 38, 45 and 65 m, of Savannah River
(SRP-I/II)~\cite{SRP} at 18 and 24 m, of
Krasnoyarsk-I/III~\cite{krasnoyarsk} at 33, 92 and 57 m,
ROVNO88-1I/2I/1S/3S~\cite{rovno88} at 18 m, ROVNO88-2S~\cite{rovno88}
at 25 m and ROVNO91~\cite{rovno} at 18 m.

Our simulation follows closely the one described in
Ref.~\cite{reactor-anomaly}. 
We use the isotopic compositions and new rates provide 
in Tab. II of Ref.~\cite{reactor-anomaly}, as well as the $\chi^2$ 
function with the covariance matrix defined in this reference.
Regarding the covariance matrix, it is important to highlight that each 
element should be multiplied by the respective rate. 
In other words, following the notation of 
\cite{reactor-anomaly}, 
each element of the covariance matrix $W$ is defined as
$W_{ij} = \sigma_{ij}^2 R_i\, R_j$, where $ \sigma_{ij}^2$ is the
correlated error between experiments $i$ and $j$ when $i\neq j$ or
simply the corresponding experiment error for diagonal elements, and
$R_i$ is the ratio of observed over expected number of events of the
experiment $i$. To obtain the theoretical rates with LED, we used the
experimental results available in
Refs.~\cite{bugey4,ill,bugey3,gosgen,rovno,rovno88,krasnoyarsk,SRP}
and the parameterization given in \cite{reactor-anomaly} to calculate
the expected reactor fluxes. We implemented all experiments using a
modified version of GLoBES \cite{globes}.

We have fitted the new rates in the LED scenario and obtained the 
allowed regions for the LED parameters $m_0$ and $a$. 
In the middle panels of Fig.~\ref{fig:allowed-regions}
we show these regions  for NH (upper panel) and IH (lower panel). 
We observe that these regions are more compatible with the limits 
coming from other oscillation experiments~\cite{mnz2010}, indicated by the 
black solid curve,  than the ones obtained by Ga data shown in the left panels 
of Fig.~\ref{fig:allowed-regions}.

\subsection{Combined Analysis}
\label{sec:combined-analysis}
Finally we show the results of the combined LED analysis for 
GALLEX and SAGE source experiments with the one for 
the 19 short baseline reactor experiments. 
In the right panels of Fig.~\ref{fig:allowed-regions} 
we show the allowed regions 
for NH (upper panel) and IH (lower panel) 
in the plane of $m_0$ and $a$ 
obtained by combining Ga source experiment 
and short baseline reactor experiments. 
We found that the combined data favor 
the nonzero value of the large extra dimension,  
2.9 $\sigma$ away from $a=0$. 

We have further combined results of 
these Ga source and short baseline reactor data and 
the data coming from KamLAND, CHOOZ and MINOS previously 
considered in Ref.~\cite{mnz2010} 
but we do not show the plot here as it is 
quite similar to what have been shown in the right panels
of Fig.~\ref{fig:allowed-regions}. 
The reason is that the region favored by 
gallium and reactor antineutrino anomalies 
and the region excluded by 
KamLAND, CHOOZ and MINOS overlap scarcely.

\section{ Discussions and Conclusions}

Current neutrino data exhibit three anomalies, 
one in the appearance mode, $\bar{\nu}_e$ excess in LSND~\cite{LSND} 
and MiniBooNE~\cite{MiniBooNE} experiments, the other two are 
deficit of $\nu_e$ in the gallium solar neutrino calibration 
experiments~\cite{gallex,sage} and of $\bar{\nu}_e$ in 
the short baseline ($ < 100$ m) reactor experiments~\cite{reactor-anomaly}. 
Possible solution to these problems involving 
oscillation into one or two species of sterile neutrinos whose 
mass squared differences are separated from 
the active ones by $\sim$ eV$^2$, have been proposed. 

In this work we show that the two of these anomalies in the
disappearance mode can be explained by an alternative solution,
oscillation of $\nu_e$ and $\bar{\nu}_e$ into sterile Kaluza-Klein
neutrinos which are present in a model with large extra dimensions
with a dimension size of $\lsim 0.6\ \mu$m, and compatible with the
limits coming from other oscillation experiments analysed in
Ref.~\cite{mnz2010}.

Let us make some comments on LED limits coming from other 
sources/considerations besides KK bulk neutrinos. 
First, cosmological and astrophysical bounds on LED (or equivalently
on the fundamental scale of gravity) due to the over production
and/or decays of KK gravitons into SM particles in various
cosmological/astrophysical environments give, in general, much
stronger bounds than the ones coming from laboratory experiments~\cite{Hall:1999mk,Hannestad:2001nq,Hannestad:2003yd,Casse:2003pj,Hannestad:2004px}.
However, since these bounds are not completely model independent and
not coming directly from the presence of the KK neutrinos, we do not
try to make a direct comparison here.

Instead, we prefer to quote some cosmological limits coming
directly from the presence of the KK neutrinos obtained 
in Refs.~\cite{Abazajian:2000hw,Goh:2001uc}.
In Ref.~\cite{Abazajian:2000hw} for the case where 
the ``normalcy'' temperature of the universe 
(considered as the temperature at which the universe should be 
free from the KK modes for graviton production, 
see the last reference in ~\cite{ADD})
was assumed to be $\lsim $ 1 GeV, 
for $\delta = 4$ ($\delta$ being the number 
of large extra dimensions of equal size), 
by requiring that neutrinos should not contribute too much 
to the energy density of the universe, 
a size larger than $\sim 1\ \mu$m 
for $m_i$ larger than 0.01 eV 
is excluded (for $\delta = 5,6$ the bounds become stronger, 
see Fig. 2 of \cite{Abazajian:2000hw}). 
This may seem to exclude our solution, but 
we can not make a direct comparison since we assumed here 
that only one, the largest extra dimension (the other 
dimensions having negligible size), can contribute 
to alter significantly the oscillation probability.

On the other hand, a complementary analysis to Ref.~\cite{Abazajian:2000hw}
was performed in Ref.~\cite{Goh:2001uc} where a bound on the size of the 
largest extra dimension was derived such that the KK modes would not cause 
any conflict between the successful theoretical predictions of the Big Bang 
Nucleosynthesis (BBN) and its observations. 
Since in this case it was assumed that only a single KK tower would 
contribute to modify BBN, we can make a direct comparison. 
From Fig. 1 of Ref.~\cite{Goh:2001uc}, we observe that 
the typical solution we found, $a \sim $ 
a few $\times\ 0.1\ \mu$m and $m_i \sim  O(0.1)$ eV is still allowed.

We note that the $\bar{\nu}_e$ excess observed in the LSND and
MiniBooNE experiments cannot be explained by the simple LED scenario
we consider in this work. In order to do that this scenario would
have to be extended (see Ref.~\cite{Davoudiasl:2002fq}), however  
the LSND/MiniBooNE anomaly is becoming much weaker with new data.

While the future MINOS and Double CHOOZ data can improve somewhat the
limits in the small $m_0$ parameter region~\cite{Machado-Planck2011},
it seems not easy to exclude or confirm the LED solution discussed in
this work.  This also seems to apply to the sterile neutrino
explanations discussed in
Refs.~\cite{reactor-anomaly,Kopp:2011qd,Giunti:2011gz}.  In fact as
far as the gallium and reactor antineutrino anomalies are concerned,
behavior of these two solutions are similar (as both of these
solutions exhibit rapid oscillations) so that it would not be so easy
to distinguish them.

Possibly, a large liquid scintillator detector with very low
background such as KamLAND~\cite{kl} using a PBq scale radioactive
source deployed in its center, capable of very good vertex
reconstruction, as discussed in
Ref.~\cite{reactor-anomaly,Cribier:2011fv}, could allow us to observe
the rapid oscillation patterns which may help in identifying the
solution to the gallium and reactor antineutrino anomalies. See also
Ref.~\cite{Formaggio:2011jt}.

\vspace{0.3cm}
\begin{acknowledgments} 
  \vspace{-0.3cm} This work is supported by Funda\c{c}\~ao de Amparo
  \`a Pesquisa do Estado de S\~ao Paulo (FAPESP), Funda\c{c}\~ao de
  Amparo \`a Pesquisa do Estado do Rio de Janeiro (FAPERJ) and
  Conselho Nacional de Ci\^encia e Tecnologia (CNPq). PANM has also
  been supported by a European Commission ESR Fellowship under the
  contract PITN-GA-2009-237920.

We would like to thank Thierry Lasserre, Maximilien Fechner and
Guillaume Mention for clarifying some details of the correlation
matrix used in the analysis done in Ref.~\cite{reactor-anomaly}, Jenny
Thomas and Robert Plunkett for useful correspondence about MINOS+
project, and Joachim Kopp for clarifications about 3+1 and 3+2 global
fits.  Three of us (PANM, HN and RZF) also would like to acknowledge
the Fermilab Theory Group for its hospitality during the last stage of
this work.

\end{acknowledgments}


\begin{thebibliography}{99}
\bibitem{footnote1} 
Regarding the number of free parameters that affects 
the oscillation physics, generic 3+1 and 3+2 models 
have, respectively, 12 and 20 free parameters, 
whereas the LED model discussed here has 8, 
only 2 more additional parameters compared to 
the standard three flavor scheme.
\end{thebibliography}
\end{document}